\documentclass[final,3p,times,twocolumn]{elsarticle}

\usepackage{hyperref}
\usepackage{amsmath}
\usepackage{xspace}
\usepackage{ulem}

\usepackage[pdftex]{color}
\usepackage{multirow}

\journal{Journal of \LaTeX\ Templates}

\begin{document}

\begin{frontmatter}

\title{A detailed study on spectroscopic performance of SOI pixel detector with a pinned depleted diode structure for X-ray astronomy}

\author[Miyazakiaddress]{Masataka Yukumoto\corref{mycorrespondingauthor}}
\cortext[mycorrespondingauthor]{Corresponding author}
\ead{yukumoto@astro.miyazaki-u.ac.jp}

\author[Miyazakiaddress]{Koji Mori}
\ead{mori@astro.miyazaki-u.ac.jp}

\author[Miyazakiaddress]{Ayaki Takeda}
\author[Miyazakiaddress]{Yusuke Nishioka}
\author[Miyazakiaddress]{Miraku Kimura}
\author[Miyazakiaddress]{Yuta Fuchita}
\author[Miyazakiaddress]{Taiga Yoshida}

\author[Kyotoaddress]{Takeshi G. Tsuru}

\author[DSaddress]{Ikuo Kurachi}

\author[Tokyoaddress]{Kouichi Hagino}

\author[KEKaddress]{Yasuo Arai}

\author[TUSaddress]{Takayoshi Kohmura}

\author[Konanaddress]{Takaaki Tanaka}

\author[Kinkiaddress]{Kumiko K. Nobukawa}


\address[Miyazakiaddress]{Department of Applied Physics, Faculty of Engineering, University of Miyazaki, 1-1 Gakuen-Kibanadai-Nishi, Miyazaki, 889-2192, Japan}
\address[Kyotoaddress]{Department of Physics, Faculty of Science, Kyoto University, Kitashirakawa Oiwake-cho, Sakyo-ku, Kyoto 606-8502, Japan}
\address[DSaddress]{D$\&$S Inc., 774-3-213 Higashiasakawacho, Hachioji, Tokyo 193-0834, Japan}
\address[Tokyoaddress]{Department of Physics, University of Tokyo, 7-3-1 Hongo, Bunkyo, Tokyo 113-0033, Japan}
\address[KEKaddress]{Accelerator Laboratory, High Energy Accelerator Research Organization (KEK), 1-1 Oho, Tsukuba 305-0801, Japan}
\address[TUSaddress]{Department of Physics and Astronomy, Faculty of Science and Technology, Tokyo University of Science, 2641 Yamazaki, Noda, Chiba 278-8510, Japan}
\address[Konanaddress]{Department of Physics, Konan University, 8-9-1 Okamoto, Higashinada, Kobe, Hyogo 658-8501, Japan}
\address[Kinkiaddress]{Faculty of Science and Engineering, Kindai University, 3-4-1 Kowakae, Higashi-Osaka, Osaka 577-8502, Japan}

\begin{abstract}
We have been developing silicon-on-insulator (SOI) pixel detectors with a pinned depleted diode (PDD) structure, named ``XRPIX'', for X-ray astronomy. In our previous study, we successfully optimized the design of the PDD structure, achieving
both the suppression of large leakage current and satisfactory X-ray spectroscopic
performance. Here, we report a detailed study on the X-ray spectroscopic
performance of the XRPIX with the optimized PDD structure. The data were obtained at
$-60$~$^\circ$C with the ``event-driven readout mode'', in which only a triggering
pixel and its surroundings are read out. The energy resolutions in full width at half
maximum at 6.4~keV are $178\pm1$~eV and $291\pm1$~eV for single-pixel and all-pixel
event spectra, respectively. The all-pixel events include charge-sharing pixel
events as well as the single-pixel events. These values are the best achieved in
the history of our development. We argue that the gain non-linearity in the low
energy side due to excessive charge injection to the charge-sensitive amplifier is a
major factor to limit the current spectroscopic performance. Optimization of the
amount of the charge injection is expected to lead to further improvement in the
spectroscopic performance of XRPIX, especially for the all-pixel event spectrum.

\end{abstract}

\begin{keyword}
X-ray detectors \sep X-ray SOIPIX \sep Monolithic active pixel sensors \sep Silicon on insulator technology
\end{keyword}

\end{frontmatter}

\section{Introduction}
\label{sec:intro}

Charge-coupled devices (CCDs) had been the most widely used image sensor both in
commercial and astronomical uses for long time since its appearance. At least
in commercial use, it has been a while since complementary metal-oxide-semiconductor
(CMOS) sensors almost completely replaced CCDs. CMOS sensors inherently possess several advantages over
CCDs such as lower power consumption, lower voltage operation, faster readout,
on-chip functionality, and so on \cite{ref:bigas}. Their lower power consumption and
lower voltage operation are especially beneficial in the consumer electronics market
and have been the driving force behind overturning the prevalence of CCDs. A similar
trend is coming in the astronomical field \cite{ref:alarcon}, and X-ray astronomy is
also no exception. Progress in X-ray astronomy over the last three decades
undoubtedly owes much to CCD detectors \cite{ref:Burke, ref:Garmire, ref:Turner,
ref:Koyama}. The X-ray CCD detectors keep advancing and new X-ray observatories
still utilize their utility and reliability \cite{ref:Predehl, ref:Tanaka, ref:Nakajima,
ref:Mori, ref:Bautz}. Under these circumstances, X-ray CMOS detectors are certainly
presenting new possibilities beyond X-ray CCDs for future X-ray astronomy.
We summarize the performance of modern X-ray CCDs and CMOS detector for X-ray astronomy in Table~\ref{tab:comp_spec}

\begin{figure*}[htbp]
  \makeatletter
   \def\@captype{table}
   \makeatother
\begin{center}
  \caption{Performance of modern X-ray CCDs and CMOS detector for X-ray astronomy.}
  \label{tab:comp_spec}
\begin{tabular}{cccc}
\hline\hline
 & pnCCD \cite{ref:Meidinger} & fast-readout CCD  & scientific CMOS \\
 & & for AXIS \cite{ref:Eric} & for Einstein Probe \cite{ref:Qinyu}\\
\hline
Number of pixels & 384~$\times$~384 & 1440~$\times$~1440 & 4096~$\times$~4096 \\
Pixel size & 75~{\textmu}m square & 24~{\textmu}m square & 15~{\textmu}m square \\
Sensor layer thickness & 450~{\textmu}m & 100~{\textmu}m & 10~{\textmu}m\\
Time resolution / Frame rate & 50~msec & $\leq$5~fps & 20~fps\\
Energy resolution at 6~keV & \multirow{2}{*}{$\sim 133$~eV at $-85$~$^\circ$C} & \multirow{2}{*}{129~eV at $-87$~$^\circ$C} & \multirow{2}{*}{140~eV at room temp.}\\
(Single-pixel event) & & &\\
\hline
\end{tabular}
\end{center}
\end{figure*}

One concern in applying conventional CMOS sensors for X-ray astronomy is its
low quantum efficiency for X-ray. Their sensor layer thickness is typically a few
{\textmu}m. A thick sensor layer of $>$100~{\textmu}m is necessary for X-ray detection up to 10~keV with
high quantum efficiency. To overcome this situation, we have been developing a novel
CMOS pixel detector, named ``XRPIX'', based on silicon-on-insulator (SOI) CMOS
technology \cite{ref:Tsuru}. The SOI CMOS technology allows both a thick
high-resistivity silicon for sensor layer and a low-resistivity silicon for
high-speed circuit to coexist in a single monolithic detector \cite{ref:Arai}. XRPIX
achieves a more than 300~{\textmu}m-thick fully depleted sensor layer.

Utilizing the on-chip functionality, XRPIX incorporates a self-trigger function in
each pixel \cite{ref:Takeda_1b}. Restricting readout pixels to a triggering and
its surroundings ones enables XRPIX to achieve a fine time resolution of
$\sim$10~{\textmu}sec with a high throughput of $\sim$1~kHz. With this
``event-driven readout mode'', XRPIX can significantly reduce non-X-ray background by
applying the anti-coincidence technique and mitigate the degree of event pile-up
for point sources.

The event-driven readout mode of XRPIX is certainly unique function beyond X-ray
CCDs. However, with regard to spectroscopic performance, XRPIX still falls behind X-ray CCDs.
We have recently developed XRPIX8 to improve the spectroscopic
performance of XRPIX introducing a Pinned Depleted Diode (PDD) structure in the
sensor layer \cite{ref:Kamehama}. The PDD structure is relatively complex compared
to our previous XRPIXs and optimization studies were necessary to finalize the
design of the PDD structure. We successfully found out the best design, achieving
both the suppression of large leakage current and satisfactory X-ray spectroscopic
performance \cite{ref:Yukumoto}. Here, we report a detailed study on the
spectroscopic performance of XRPIX8 with the optimized PDD
structure\footnote{XRPIX8 has five subversions. ``XRPIX8 with the optimized
PDD structure'' here is specifically XRPIX8.5 in the terminology of
\citep{ref:Yukumoto}.}. We also provide our post data-acquisition (DAQ)
processing. The data was obtained with the event-driven readout mode at
$-60$~$^\circ$C, the lowest operation temperature of our intended use.
This spectroscopic performance study was performed for not only the
single-pixel events but also the all-pixel events including charge-sharing pixel
events. Evaluation only with single pixel events is useful for understanding basic
response of pixel detectors, especially in the early stage of the
development. Actually, we have been evaluating our XRPIX with single pixel
events. On the other hand, the fraction of single pixel events is typically less
than half of the total, about 20~\% in our case. Therefore, evaluation with
all-pixel events including charge-sharing pixel events is necessary to utilize all
the events detected.

Section~2 gives the device description of XRPIX8 and data format of the
event-driven readout mode. Section~3 describes the post DAQ processing step by step.
Section~4 presents the details of the spectroscopic performance of XRPIX8. Then,
Section~5 discusses the causes limiting the current spectroscopic performance.
Finally, Section 6 summarizes this study.

\section{Device and data description}
\label{sec:device}

\begin{figure}[htbp]
  \makeatletter
   \def\@captype{table}
   \makeatother
\begin{center}
  \caption{Specifications of XRPIX8.}
  \label{tab:chip_spec}
\begin{tabular}{ccc}
\hline\hline
Chip size  & 6~mm~square\\
Number of pixels & 96~$\times$~96\\
Pixel size  & 36~{\textmu}m~square\\
Sensor layer thickness  & 300~{\textmu}m\\
\hline
\end{tabular}
\end{center}
\end{figure}

The details of XRPIX8, including the PDD structure, are given in \cite{ref:Yukumoto}. Table~\ref{tab:chip_spec} particularly summarizes specifications of XRPIX8 relevant to this study. As noted in the previous section, the large sensor layer thickness is one of the most important features of XRPIX and is advantageous for X-ray detection in the higher energy band even above 10~keV. Since the pixel size is kept as small as possible for high spatial resolution, the thickness of the sensor layer is large relative to the pixel size in comparison with other X-ray silicon pixel detectors. We apply a high back bias voltage of $-300$~V to the sensor layer in order to suppress the overspreading of charge cloud and make conventional grade selection applicable \cite{ref:Hagino_3b}. The remaining description in this section focuses on the pixel circuit and the format of data, which are related to the following post-DAQ processing and spectroscopic performance study.

\begin{figure}[htbp]
        \centering
        \includegraphics[scale=0.4]{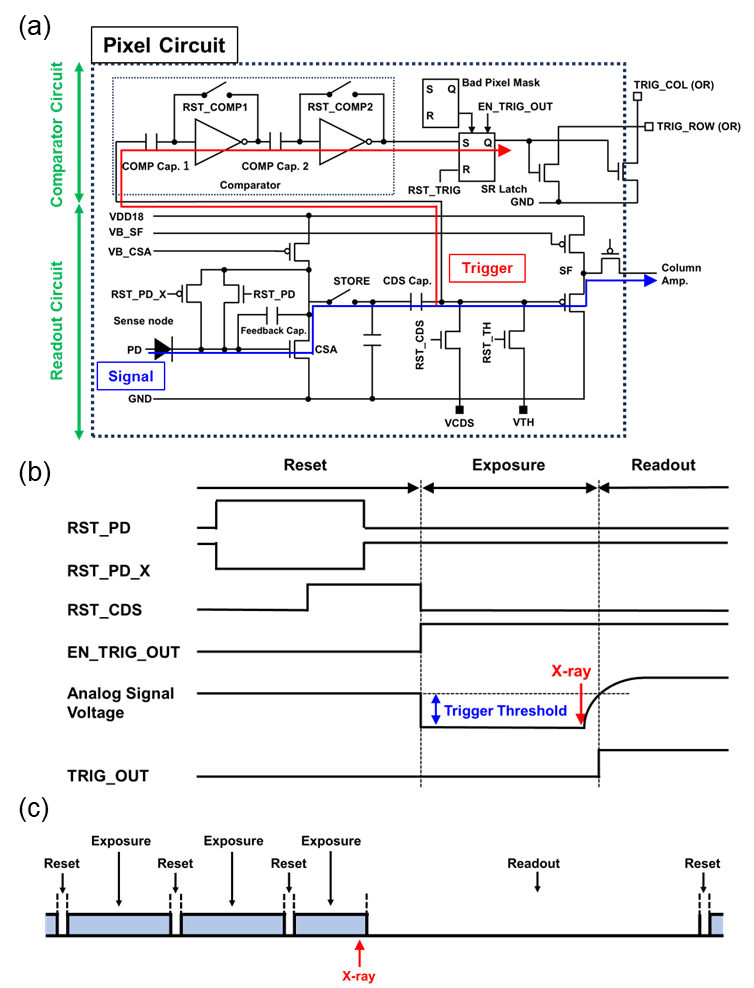}
        \caption{(a) Pixel circuit of XRPIX8. (b) Timing diagram of the operation.
        (c) A part of the time sequence of operation. The blue periods correspond to live time periods.}
        \label{fig:pix_circuit}
\end{figure}

Fig.~\ref{fig:pix_circuit} (a) shows the pixel circuit of XRPIX8.
The pixel circuit consists of a readout circuit and a comparator circuit. The
readout circuit includes a charge-sensitive amplifier (CSA) circuit to convert the
charge produced by an X-ray photon to a signal voltage and a correlated double sampling (CDS)
circuit to suppress the kT/C reset noise in the sense node. The comparator circuit
is for trigger output.

Fig.~\ref{fig:pix_circuit} (b) shows the timing diagram of the operation. Readout circuits are reset by RST\_PD, RST\_PD\_X and RST\_CDS. An exposure starts when RST\_CDS is turned off and EN\_TRIG\_OUT is turned on. EN\_TRIG\_OUT makes comparator circuits ready to output a digital trigger signal, TRIG\_OUT. Charge accumulation from the sense node starts at this point. When the signal voltage exceeds the trigger threshold at a certain pixel, TRIG\_OUT is output from the comparator circuit of the pixel. An exposure time is defined as the time from when the RST\_CDS is turned off until the digital trigger signal is output, and hence each event has a different exposure time. A more detailed timing diagram for the pixel circuit and description are given in \cite{ref:Takeda_1b}.

Fig.~\ref{fig:pix_circuit} (c) shows a part of the time sequence of operation. We reset all the pixel circuits every 100~{\textmu}s unless an X-ray is detected and a digital trigger signal is output during the period. It takes $\sim$~30~{\textmu}s to reset. The period of 100~{\textmu}s can be modified according to an observation purpose. When the digital trigger signal is output due to incident X-ray, the analog signal voltages of 8~$\times$~8~pixels centered on the triggering pixel are read out and stored as an event. It takes $\sim$~310~{\textmu}s to readout. Therefore, the live time fraction of an observation depends on its incident X-ray count rate.

If the readout size is restricted to be smaller than the 8~$\times$~8~pixel region, the readout time will be correspondingly shorter. However, the readout size should be sufficiently larger than the charge cloud sizes and should be processed with $2^{n}$ from a design perspective. The 8~$\times$~8~pixel region is the smallest size that satisfies the above.

\begin{figure}[htbp]
        \centering
        \includegraphics[scale=0.5]{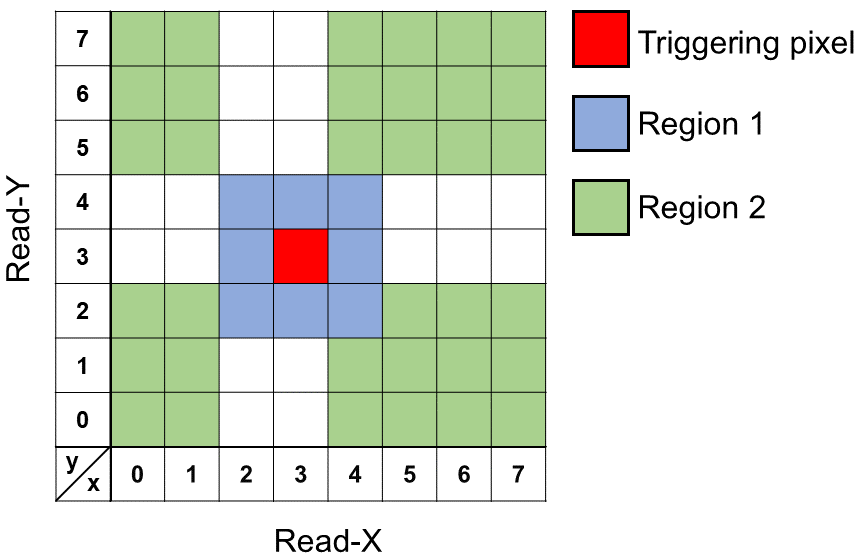}
        \caption{8~$\times$~8~pixel map of an event in the relative coordinate.}
        \label{fig:data_format}
\end{figure}

Two coordinate systems are used
here. One is the absolute coordinate to give the position in the
96~$\times$~96~pixels, in which ``column address'' and ``row address'' are used to
express horizontal and vertical axes, respectively. The other is the relative coordinate to
give the position in the analog-signal-readout  8~$\times$~8~pixels, in
which ``read-X'' and ``read-Y'' are used to express horizontal and vertical axes,
respectively. Fig.~\ref{fig:data_format} shows the pixel map of an event in the
relative coordinate. The triggering pixel is placed at (read-X, read-Y) = (3, 3).
Charge sharing due to the charge diffusion during the drift in the sensor layer
commonly occurs in silicon pixel sensors. The surrounding 8~pixels, labeled as
region~1, could have the signal charge by the charge sharing considering
the pixel and charge cloud sizes \cite{ref:Hagino_3b}. Columns at read-X =
2, 3 and rows at read-Y = 3, 4 can be affected by the trigger function interference
(see next section). Outer pixels, labeled as region 2, are not expected to share the
signal charge nor to be affected by the trigger function interference. The
analog signal voltages of 64 pixels are digitized by an analog-to-digital converter and stored as raw values. The unit of the raw value is analog to digital units (ADU), which represents the digitized analog signal voltage, and 1~ADU is 488~{\textmu}V.

\section{Experiments and data processing}\label{sec:data_proc}

In this experiment, XRPIX8 was irradiated from the back side with
X-rays/$\gamma$-rays from the radioisotopes, $^{57}$Co and $^{241}$Am, and an X-ray
tube with a Ti target. A back bias voltage of $-300$~V was applied to fully deplete
the sensor layer.
We acquired two data sets, ``evaluation data'' and ``calibration
data''. The evaluation data is the subject of this spectroscopic
performance study, and final numbers are deduced from the evaluation
data. If the evaluation data is calibrated with themselves, the
results would be biased to be better than they actually
are. Therefore, we derived the calibration parameters from the
calibration data and then calibrated the evaluation data with them in
the gain correction and energy calibration (step~4 and 5 in the
following).
The evaluation data was taken from the 8~$\times$~8~pixel region with the column
addresses of 17--24 and the row addresses of 20--27.
The calibration data was taken from the 16~$\times$~16~pixel region centered on the 8~$\times$~8~pixel region. This
is because an event whose triggering pixel is located at the outer pixels of the
8~$\times$~8~pixel region has information of pixels outside that region.

\begin{figure}[htbp]
        \centering
        \includegraphics[scale=0.4]{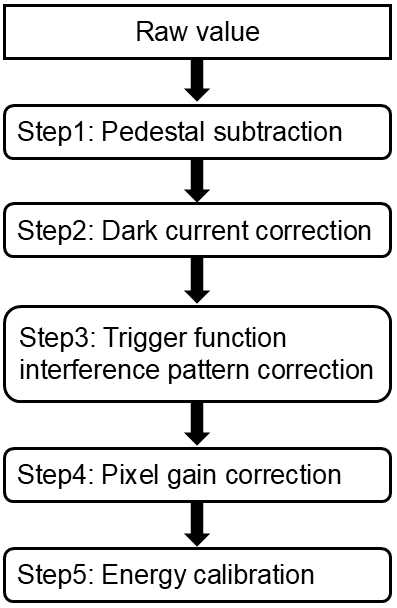}
        \caption{Flowchart of post DAQ processing.}
        \label{fig:ana_step}
\end{figure}

Fig.~\ref{fig:ana_step} shows a flowchart of post DAQ processing. There are
five steps in creating X-ray spectra from raw X-ray events. From step~1 to step~3,
the same data processing applies to both calibration and evaluation data
independently. From step~4 to step~5, parameters derived from the calibration data
are used to calibrate the evaluation data. The details of each step are described in
the following. The data used in the figures shown in the following subsections are
the evaluation data otherwise noted.

\subsection*{Step 1: Pedestal subtraction}

\begin{figure}[htbp]
        \centering
        \includegraphics[scale=0.39]{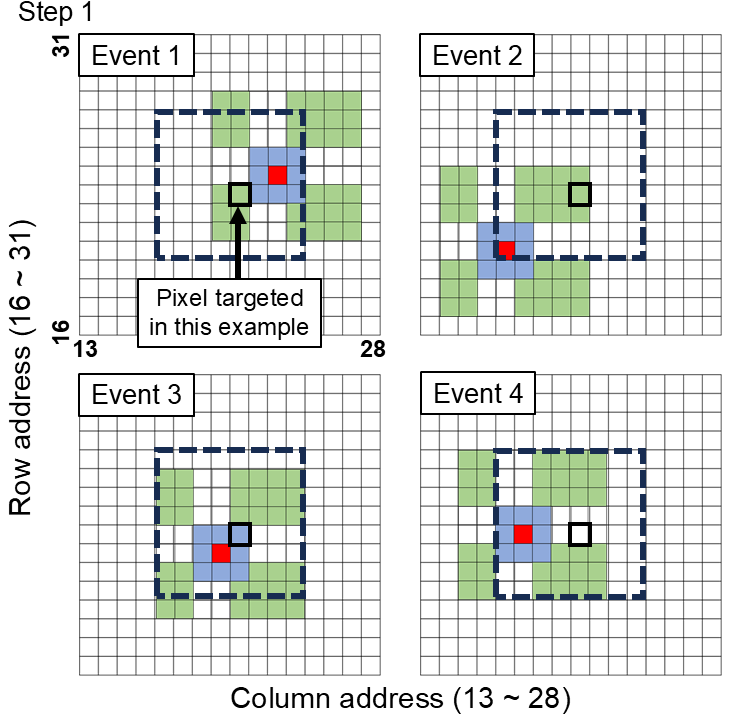}
        \caption{Illustration of the pedestal calculation of a given pixel,
        which is edged with a black solid line in this figure. The black dashed line
        indicates the 8~$\times$~8 region where the spectroscopic performance is
        investigated. The pixel targeted in this example is contained in the region~2 of the relative coordinate in
        the cases of event~1 and event~2, but not in the cases of event~3 and
        event~4. In this example, the raw values of that pixel in the event~1 and
        event~2 are used for the pedestal calculation of that pixel.}
        \label{fig:pedstal_calc}
\end{figure}

\begin{figure}[htbp]
        \centering
        \includegraphics[scale=0.36]{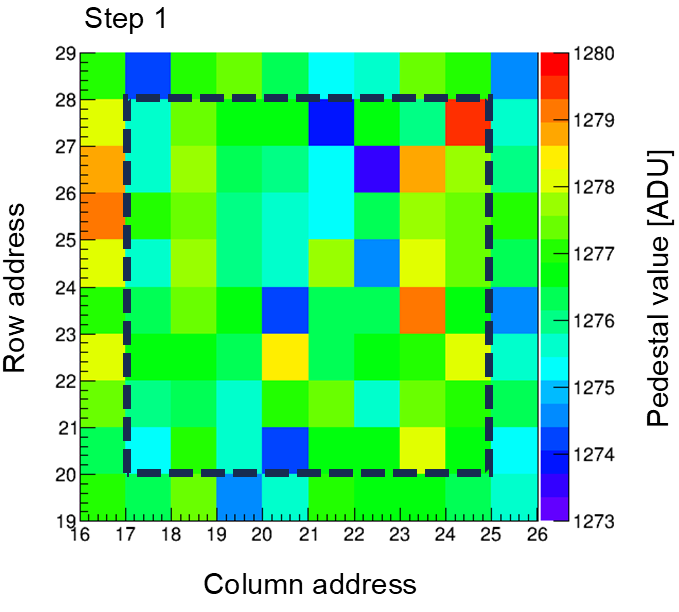}
        \caption{Pedestal value map of the 10~$\times$~10~pixels. The black dashed
        line indicates the 8~$\times$~8 region where the spectroscopic performance
        is investigated}
        \label{fig:pedstal_map}
\end{figure}

Each pixel has a different pedestal value in the absolute coordinate.
Fig.~\ref{fig:pedstal_calc} illustrates the calculation procedure of the pedestal
value of a given pixel. All events containing that pixel in the region~2 of
the relative coordinate are extracted, and the average of the raw values of that
pixel in such events is calculated and used as the pedestal value of that pixel. The
pedestal value for each single pixel is calculated in this way.
Fig.~\ref{fig:pedstal_map} shows the resulting pedestal value map. The pedestal
values are calculated for the pixels outside the 8~$\times$~8 region because they
could be in the region~1 of the relative coordinate due to the charge
sharing. We subtract the pedestal value from the raw value for each pixel, and the
raw value minus the pedestal value is recorded as the pulse height.

\subsection*{Step 2: Dark current correction}
\begin{figure}[htbp]
      \centering
        \includegraphics[scale=0.27]{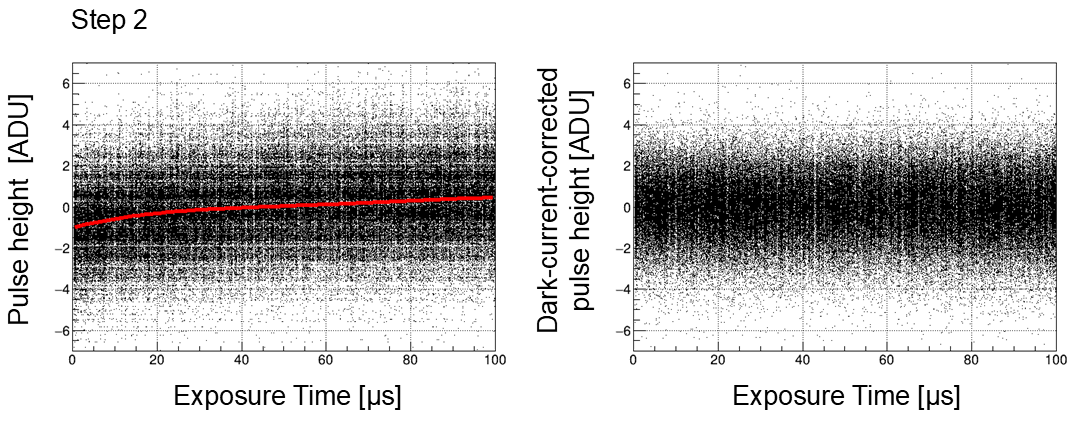}
        \caption{Pulse height (left) and dark-current-corrected pulse height (right) as a function of the exposure time in the region~2.
        The red solid line in the left panel is the best fit curve with the fourth-order polynomial function.}
        \label{fig:light_curve}
\end{figure}

Fig.~\ref{fig:light_curve} left shows the pulse height in the region~2 as a function of the exposure time. It is clear that the
pulse height correlates with the exposure time due to larger contributions of dark
current in longer exposure time events.
Assuming that the dark current is constant, the pulse height should increase linearly as a function of exposure time. This actually applies after 20~{\textmu}s, but does not before 20~{\textmu}s. The detailed mechanism is currently unknown and we empirically approximate the relation
with a fourth-order polynomial function. We then add a correction to the pulse
height pixel by pixel and event by event using the polynomial function.
Fig.~\ref{fig:light_curve} right shows the dark-current-corrected pulse height as a
function of the exposure time. The dark-current-corrected pulse heights are
apparently distributed around 0~ADU with no exposure time dependence, indicating
that the dark current contribution is successfully eliminated.

\begin{figure}[htbp]
        \centering
        \includegraphics[scale=0.38]{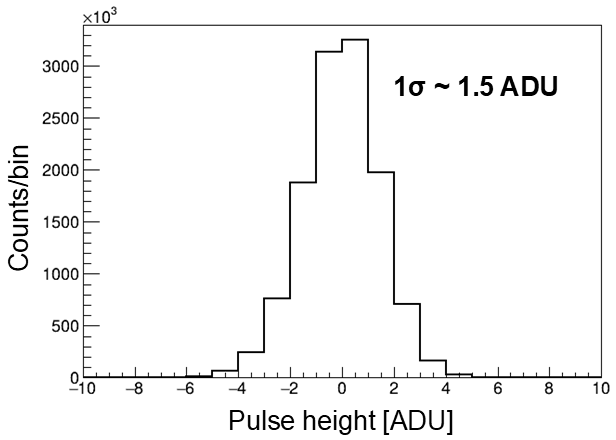}
        \caption{Pulse height distribution in the region~2 of the relative coordinate after the dark current correction.}
        \label{fig:zero_peak}
\end{figure}

Fig.~\ref{fig:zero_peak} shows the pulse height distribution in the region~2 after
the dark current correction. This plot is a projection of Fig.~\ref{fig:light_curve}
right onto the y axis. We hereafter call this pulse height distribution as the
zero-peak spectrum.

\subsection*{Step 3: Trigger function interference pattern correction}

\begin{figure}[htbp]
        \centering
        \includegraphics[scale=0.28]{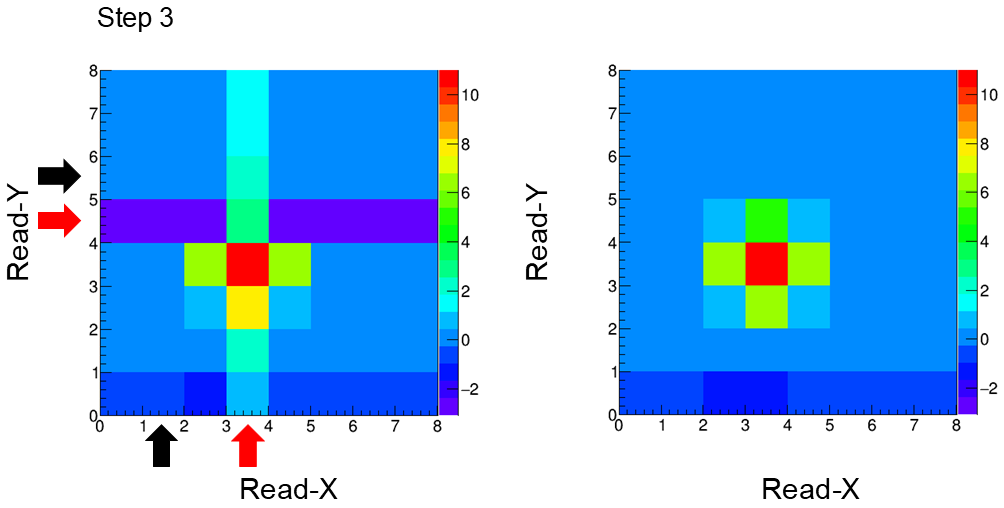}
        \caption{Map of the mean pulse height after step~2 of the event with a single trigger signal output in the relative coordinate before (left) and after (right) trigger function interference pattern correction. The characteristic pattern and the reference lines are indicated by red and black arrows in the left panel, respectively.}
        \label{fig:trigger_pattern}
\end{figure}

Fig.~\ref{fig:trigger_pattern} left shows a map of the mean pulse height after
step~2 of the events with a single trigger signal output in the relative coordinate.
A characteristic pattern appears at the column at read-X~=~3 and the row at read-Y~=~4.
In order to measure the pulse height difference in the characteristic pattern
lines, we define the column at read-X~=~1 and the row at read-Y~=~5 as reference
lines. The differences in pulse height compared to the reference lines are $+1.8\pm0.1$~ADU
and $-3.3\pm0.1$~ADU in the column of read-X~=~3 and the row of read-Y~=~4,
respectively, except for the pixels in the region~1 and triggering pixel in the relative coordinate.
As shown in Fig.~\ref{fig:trigger_pattern} right, the characteristic pattern disappears by
subtracting 1.8~ADU from the pulse heights of the pixels at read-X~=~3
and adding 3.3~ADU to the pulse heights of the pixels at read-Y~=~4.
In doing so, the pulse height of the pixel at (read-X, read-Y) = (3, 4),
where the characteristic pattern lines overlap, is added by 1.5~ADU.
After the correction, the pulse height distribution around the
triggering pixel becomes more symmetric as expected.

Regardless of whether the signal charge is read out or not, when the comparator
circuit operates, this characteristic pattern always appears. We confirmed it in the
following two experiments. In the first experiment, we fixed the central pixel coordinate of the
8~$\times$~8~pixels readout region at (X, Y) and opened the trigger mask\footnote{The trigger mask is a function that prevents the specified pixel from sending out a digital trigger signal.} only at (X, Y$+20$). In this case, only the pixel at (X, Y$+20$) could send out a digital trigger signal, and due to the trigger, the 8~$\times$~8~pixels centered at (X, Y)
were read out without no signal charge because no X-ray photon came in the
8~$\times$~8~pixels. In this experiment, the characteristic pattern appeared at the
column of read-X~=~3, but did not appear at the row of
read-Y~=~4. When we opened the trigger mask only at (X$+20$, Y), the characteristic
pattern appeared at the row of read-Y~=~4, but did not appear at the column of read-X~=~3.
In the second experiment, we used a non-triggered frame mode, in which all pixels are read out serially without using the trigger function \cite{ref:Ryu_1}. Even in this mode, the comparator circuit operates regardless of the pixel readout sequence. We took two data sets with trigger thresholds of 20~mV and 500~mV. The former is the same as that of the event-driven readout mode, and the latter is way above signal voltages produced by incident X-rays used in this experiment. The characteristic pattern appeared in the data taken with the trigger threshold of 20~mV, but not with 500~mV. In addition to these experiments, we also note that the offset of the characteristic pattern was constant regardless of incident X-ray energy.

The characteristic pattern is likely due to interference between the digital trigger lines and the sense node/the CSA circuit as the digital trigger lines run through pixels vertically and horizontally. However, the mechanisms of why the horizontal characteristic pattern appears one-row above the triggering pixel and why the offsets in pulse height have reverses in the vertical and horizontal characteristic lines are currently unknown.

An event could have multiple trigger signals because the signal voltage of
the charge-sharing pixel sometimes exceed the trigger threshold. In this case, the 8~$\times$~8~pixels
centered on the triggering pixel with the larger column and row
addresses are read out, and other triggering pixels are consequently placed at
(read-X, read-Y) = (2, 2) and/or (2, 3) and/or (3, 2). Then the characteristic
pattern can also appear at the column of read-X~=~2 and/or the row of read-Y~=~3 due
to the non-centered triggering pixels. The correction is made depending on the
number and position of the triggering pixels.

\subsection*{Step 4: Pixel gain correction}

\begin{figure}[htbp]
      \centering
   \includegraphics[scale=0.38]{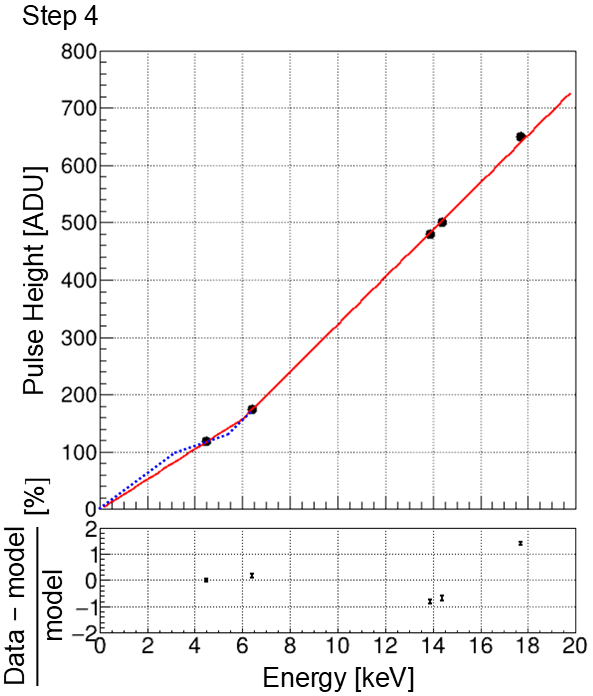}
   \caption{Center value in pulse height of major emission line as a function of energy in a specific
         pixel, which is made from the single-pixel event spectrum of the
         calibration data. The red solid line is the best fit gain curve. The blue dotted line below 6.4~keV is the putative gain curve discussed in step~5. The plot below shows the residuals between the data points and the best fit gain curve as a percentage.}
        \label{fig:calib_plot}
\end{figure}

We define a split threshold to judge how the charge-sharing occurs in an event. If
the pulse height after step~3 of a surrounding pixel in the region~1 exceeds the
split threshold, we acknowledge that the surrounding pixel contains signal
charge. Considering the charge cloud and pixel sizes, the charge-sharing can
occur within 4~pixels at the most including the triggering pixel
\cite{ref:Hagino_3b}. We then sort all the events into single-pixel
events, double-pixel events, triple-pixel events and quadruple-pixel events
depending on the number of pixels having signal charge. The split threshold is
set to be 3 times the standard deviation of the zero-peak spectrum
(Fig.~\ref{fig:zero_peak}).

With the calibration data, we first make
single-pixel event spectra pixel by pixel and measured center values of major
X-ray/$\gamma$-ray emission lines in pulse height. Fig.~\ref{fig:calib_plot} shows the
center value in pulse height of major emission line as a function of X-ray energy measured in a specific
pixel. The data points at 6.4 and 14.4~keV and those at 13.9 and 17.7~keV come from
the $^{57}$Co and $^{241}$Am spectra, respectively. The data point at 4.5~keV is
obtained from the X-ray tube spectrum with a Ti target. These data points do
not appear to follow a straight line through the origin: the slope appear to be
steeper in the higher energy side.

We here adopt a broken line model to fit the
data points. A red solid line in Fig.~\ref{fig:calib_plot} represents the best fit
gain curve in this example. In such a way, a gain curve is determined for each
8~$\times$~8 pixels with the calibration data, and the individual pulse heights of
the evaluation data are calibrated with the gain curve. The energy of an event is
calculated by summing up the calibrated pulse height of the triggering pixel and those
of its surrounding pixels above the split threshold.

\begin{figure*}[h]
      \centering
   \includegraphics[scale=0.38]{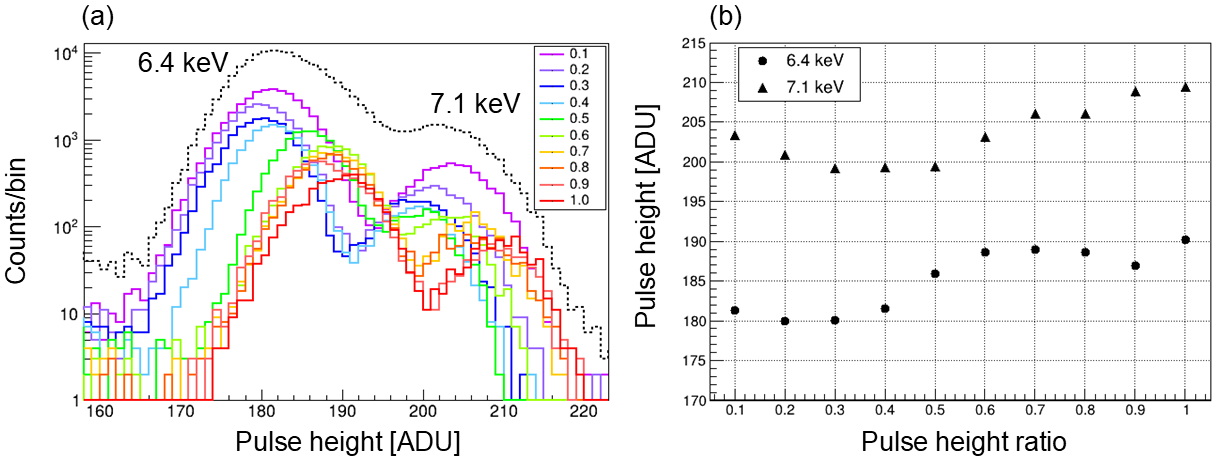}
          \caption{(a) 6.4 and 7.1~keV emission line spectra of the double-pixel events. The
          black-dotted-line spectrum is made with all the double-pixel events. The solid-line spectra are
          made by dividing the events according to the pulse height ratio of the pulse height of the adjacent charge-sharing pixel
          to that of the triggering pixel. The pulse height ratio becomes larger from cold to hot colors from 0.1 to 1.0. (b) The center values of 6.4 and 7.1~keV emission lines as a function of the pulse height ratio.}
        \label{fig:pat20_spec}
\end{figure*}

Although we did not obtain X-ray data below 4.5~keV, we realized that the
linearity between the energy and the pulse height was not maintained in the low
energy side by analyzing double-pixel event data. Since double-pixel events share the signal charge with two pixels, the charge-sharing pixel can have small charge that is equivalent to those low energy X-rays produce. Therefore, albeit indirectly, we can investigate the gain linearity in the low energy side using the double-pixel events.
Fig.~\ref{fig:pat20_spec} (a) shows the 6.4 and 7.1~keV emission line
spectra of the double-pixel events made with the evaluation data. It is obvious that
the spectrum significantly varies depending on the pulse height ratio (PHR) of the
adjacent charge-sharing pixel to that of the triggering pixel.
Fig.~\ref{fig:pat20_spec} (b) shows the center values of 6.4 and 7.1~keV emission lines as a function of
PHR. As the PHR increases, the center value of emission line shifts first to the lower energy side and
then to the higher energy side.
If the gain linearity is maintained in the low energy side, neither the spectral shape nor the center value changes regardless of the value of the PHR.
This indicates that the linear relationship does not
hold in the low energy side. We refrain from introducing a further complex
modification to the broken line model without a physical basis and leave these
distortions as systematic uncertainties.

  \begin{figure}[htbp]
          \centering
          \includegraphics[scale=0.35]{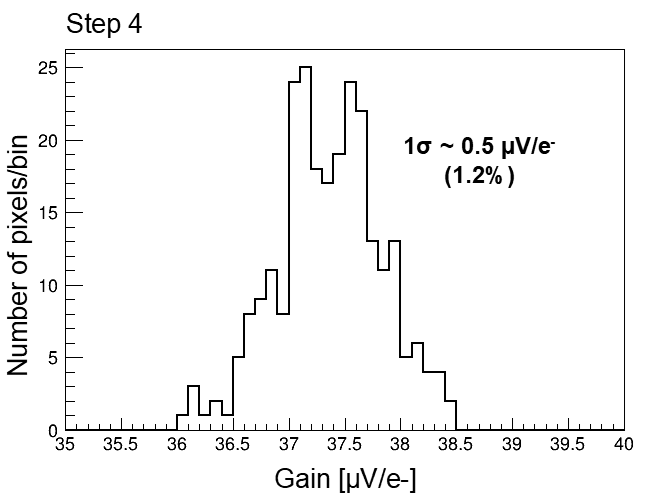}
          \caption{Histogram of the output gain value in the high energy side of the broken line model.}
          \label{fig:gain_variation}
  \end{figure}

Fig.~\ref{fig:gain_variation} shows the histogram of the output gain value in the
high energy side of the broken line model measured with the calibration
data. The output gain is calculated from the slope of the high energy side of the broken line model using
1~ADU = 488~{\textmu}V, instrumentation amplifier gain on our evaluation board of 2.0, and a
mean ionization energy in silicon of 3.65~eV \cite{ref:janesick}.
The central value is comparable with that of our study with previous XRPIX \cite{ref:Harada}. The
pixel-to-pixel gain variation is 1.2~\% in standard deviation, which is slightly larger
than 0.9~\% obtained in our previous study \cite{ref:Kodama}.

\subsection*{Step 5: Energy calibration}

\begin{figure*}[h]
      \centering
        \includegraphics[scale=0.36]{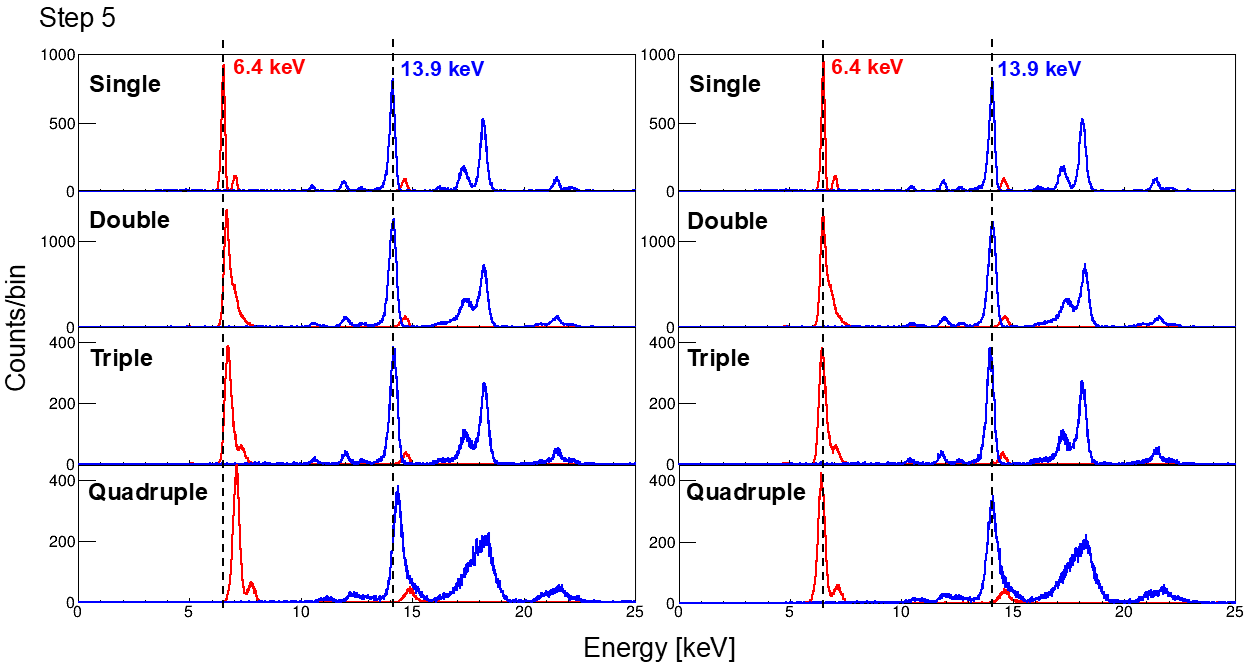}
      \caption{Single-, double-, triple- and quadruple-pixel event spectra before (left) and after (right) energy calibration. $^{57}$Co and $^{241}$Am spectra are shown in red and blue colors, respectively. The black dashed lines denote the emission line center of 6.4~keV and 13.9~keV in the spectra of the single-pixel events for eye guide.}
        \label{fig:peakshift}
\end{figure*}

Fig.~\ref{fig:peakshift} left shows the single-, double-, triple- and
quadruple-pixel event spectra from $^{57}$Co and $^{241}$Am after step~4. In spite
of the gain correction in step~4, the center energies of emission lines in the single-, double-,
triple-, and quadruple-pixel events spectra are getting higher in this order.
This systematic shift could happen if undetected charge below the split threshold is significant. Single-pixel events have the highest chance to be affected by this effect, and double-, triple-, quadruple-events have lower chance in this order. In order to test this effect, we made the spectra summing up pulse heights of the 3~$\times$~3~pixel centered on the triggering pixel and found that systematic shift shown in Fig.~\ref{fig:peakshift} left barely change. Then, this systematic shift may
suggest that approximating the gain curve with a broken line
model is too simplistic. Larger charge-sharing events tend to have more signal
pixels whose pulse height comes to the low energy side. Therefore, this result
indicates that the conversion ratio from ADU to energy in the low energy side is too
high. In other words, the slope of the broken line model in the low energy side in
Fig.~\ref{fig:calib_plot} is too low, requiring another break point for the
model. The putative gain curve discussed here is also shown in
Fig.~\ref{fig:calib_plot}. This winding gain curve is also suggested to some
extent by the behavior shown in Fig.~\ref{fig:pat20_spec}.

We re-calibrate the energy scale for each event pattern with the calibration
data spectra corresponding to Fig.~\ref{fig:peakshift} left and then apply the
derived energy scale to the evaluation data spectra. Fig.~\ref{fig:peakshift} right
shows the spectra after the final energy calibration, in which the systematic shift
of the center energies disappears.

\section{Spectroscopic performance}\label{sec:experiment}

\begin{figure*}[h]
      \centering
        \includegraphics[scale=0.42]{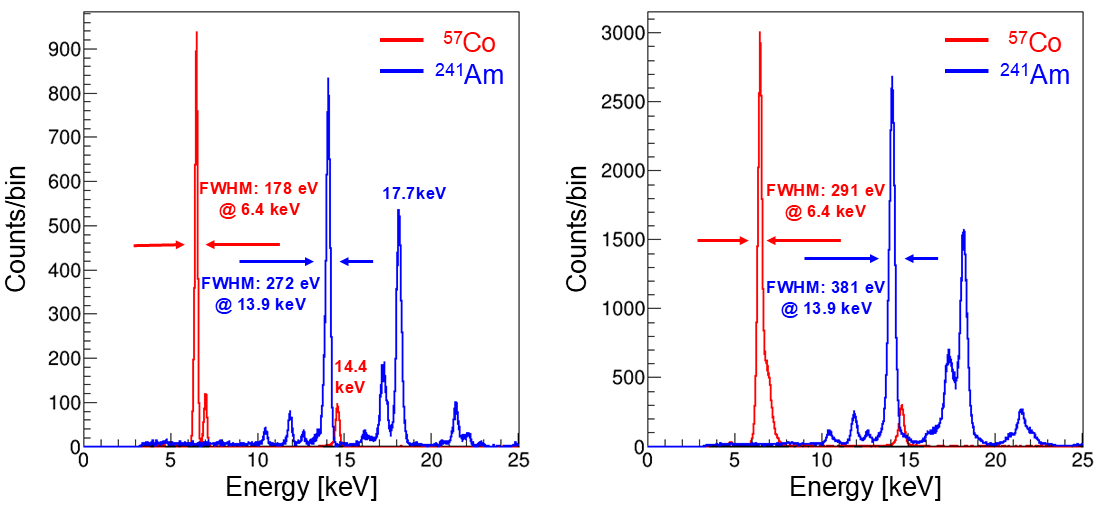}
        \caption{$^{57}$Co (red) and $^{241}$Am (blue) spectra of the single-pixel events (left) and all-pixel events (right) obtained with XRPIX8 with the optimized PDD structure.}
        \label{fig:m60d_spec}
\end{figure*}

Fig.~\ref{fig:m60d_spec} shows $^{57}$Co and $^{241}$Am spectra of the single-pixel
events and all-pixel events obtained with XRPIX8 with the optimized PDD structure.
All-pixel events include the single-, double-, triple- and quadruple-pixel events, and the all-pixel event
spectrum is simply the sum of the spectra shown in Fig.~\ref{fig:peakshift}
right. The energy resolutions at 6.4 and 13.9~keV are summarized in
Table~\ref{tab:summary}. The energy resolution at 6.4~keV of the single-pixel event
spectrum is $178\pm1$~eV in full width at half maximum (FWHM), which is the best
value ever obtained in our history of the XRPIX development and comparable to that
of CCDs used in the XRISM observatory \cite{ref:Mori2022}. On the other hand,
the energy resolution at 6.4~keV of the all-pixel event spectrum is $291\pm1$~eV in
FWHM, which is not as good as that of CCD. Quantitative evaluation of the
degradation from the single-pixel event spectrum to the all-pixel event spectrum
will be given in section~\ref{sec:discussion}.

\begin{figure}[htbp]
  \makeatletter
   \def\@captype{table}
   \makeatother
\begin{center}
  \caption{Energy resolution.}
  \label{tab:summary}
\begin{tabular}{ccc}
\hline\hline
Event pattern & 6.4~keV & 13.9~keV \\
& (FWHM) [eV]& (FWHM) [eV]\\
\hline
Single & $178\pm1$ & $272\pm2$ \\
All & $291\pm1$ & $381\pm2$\\
\hline
\end{tabular}
\end{center}
\end{figure}

\section{Discussion}
\label{sec:discussion}

Thanks to the optimized PDD structure and the refined data calibration, we achieved
the best spectroscopic performance in our development history. However,
there is still room for improvement in the spectroscopic performance of
XRPIX. Fig.~\ref{fig:pat20_spec} exhibits a clear deviation from the linear relationship between the energy and the
pulse height in the low energy side, which is a serious drawback of the current
XRPIX. We consider that the gain non-linearity in the low energy side comes from
excessive charge injection to the CSA circuit in Fig.~\ref{fig:pix_circuit}.

The CSA circuit is reset by turning on the CMOS switch that is shown as a pair
of RST\_PD (NMOS) and RST\_PD\_X (PMOS) in Fig.~\ref{fig:pix_circuit}, and then
charge injection into the CSA circuit occurs at the time when the CMOS switch is
turned off, namely the start of exposure. The amount of the charge injection is determined by the design of the
CMOS switch.
Because the thick depletion layer is one of the main characteristics
of XRPIX, the circuitry design was aimed for a wide dynamic range covering the
higher energy band up to several tens of keV. To this end, based on a circuit
simulation, the size of the PMOS transistor is designed to be significantly larger
than that of the NMOS transistor so that a large number of holes, the opposite of
electrons in the signal charge, are injected to CSA. Actually, XRPIX8 has the wide
dynamic range and the good linearity in the high energy side in the evaluation of
the single-pixel events. On the other hand, due to excessive charge injection, the
low energy side is considered to have fallen outside the linear region.
The circuit simulation with an excessive charge injection actually reproduced that the gain in the low energy side becomes low compared to that in the high energy side as we experienced.
Optimization of the amount of the charge injection is our next step to resolve the
gain non-linearity in the low energy side and to further improve the spectroscopic
performance of XRPIX.

\begin{figure}[htbp]
      \centering
        \includegraphics[scale=0.37]{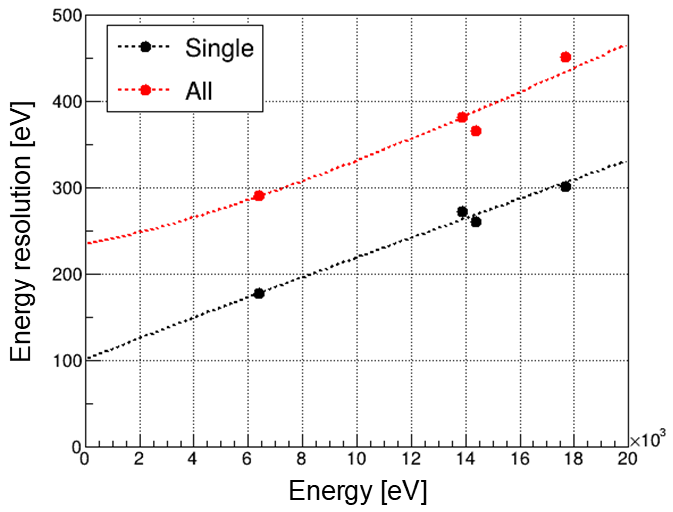}
        \caption{Energy resolution as a function of energy measured from the single-pixel event
        spectrum (black) and all-pixel event spectrum (red). The dashed lines are the
        best fit curves with Eq.~\ref{eq:Eres}.}
        \label{fig:res_func}
\end{figure}

\begin{figure}[htbp]
  \makeatletter
   \def\@captype{table}
   \makeatother
\begin{center}
  \caption{Best-fit parameters of $\sqrt{{\mathstrut (\mathop{\sigma_{\mathrm{R}}}^{2} + \mathop{\sigma_{\mathrm{D}}}^{2})} }$  and $\sigma_\mathrm{gain}$ with Eq.~\ref{eq:Eres}.}
  \label{tab:fit_param}
\begin{tabular}{ccc}
\hline\hline
Event pattern & $\sqrt{{\mathstrut (\mathop{\sigma_{\mathrm{R}}}^{2} + \mathop{\sigma_{\mathrm{D}}}^{2})} }$  [e$^{-}]$ & $\sigma_\mathrm{gain}$ [\%] \\
\hline
Single & 11.8 & 0.5 \\
All & 18.8 & 0.7\\
\hline
\end{tabular}
\end{center}
\end{figure}

The energy resolution in FWHM ($\Delta E$) is in general expressed as
\begin{equation}
  \label{eq:Eres}
  \Delta E = 2 \sqrt{2\ln 2}\, W
  \sqrt{\frac{E}{W}F +
  n^{*} \left( \sigma^{2}_{\mathrm{R}} + \sigma^{2}_{\mathrm{D}} \right) +
  \left( \frac{E}{W}\sigma_{\mathrm{gain}} \right)^2},
\end{equation}
where $E, W, F, n^{*}, \sqrt{{\mathstrut (\mathop{\sigma_{\mathrm{R}}}^{2} + \mathop{\sigma_{\mathrm{D}}}^{2})} }$ and $\sigma_\mathrm{gain}$
are the incident X-ray energy, mean ionization energy in silicon of 3.65~eV, Fano
factor of 0.128 \cite{ref:Kotov}, effective number of pixels
involved in the calculation of the energy, energy-independent noise per pixel, and pixel gain variation,
respectively \cite{ref:Kodama}. The first term in the square root is the Fano noise.
$\sigma_\mathrm{R}$ and $\sigma_\mathrm{D}$ in the second term are
the readout noise of the XRPIX electronics and dark current noise.
Assuming that these noises operate independently pixel by pixel, the
contribution of the collective noise per pixel is multiplied by $n^{*}$. In the case
of the single-pixel event spectrum, $n^{*}$ is unity by definition. Whereas, in the
case of the all-pixel event spectrum, $n^{*}$ depends on $E$ because the charge
cloud size depends on $E$ \cite{ref:Hagino_3b}. We measured the number ratio
of each event pattern using major emission lines in the all-pixel event spectrum and
estimated the energy dependency of $n^{*}$. Then, we derived an empirical relation
between $n^{*}$ and $E$ as $n^{*}=0.9\times 10^{-5} E + 2.1$, where $E$ is in the
unit of eV. The third term in the square root would be ideally zero because we
corrected the pixel gain variation, but realistically has a certain value due to
imperfection of our correction.

Fig.~\ref{fig:res_func} plots $\Delta E$ as a function of energy measured from the single-pixel event
spectrum and all-pixel event spectrum. Well-isolated emission lines at 6.4, 13.9, 14.4 and 17.7~keV were selected
in this evaluation. The best fit curves with Eq.~\ref{eq:Eres} are also shown and
parameters obtained are summarized in Table~\ref{tab:fit_param}. The values of
$\sqrt{{\mathstrut (\mathop{\sigma_{\mathrm{R}}}^{2} + \mathop{\sigma_{\mathrm{D}}}^{2})} }$ are obtained as 11.8~e$^{-}$ and 18.8~e$^{-}$ from the single- and all-pixel event spectra, respectively. The former is fully consistent
with the value from our previous study, in which single-pixel events from one
specific pixel were used \cite{ref:Yukumoto}. The latter is significantly larger
than the former. The cause of this degradation in the all-pixel event spectrum is
likely the gain non-linearity in the low energy side. Even limiting the evaluation
energy to the high energy side, the multi-pixel event spectrum becomes distorted if
the gain linearity was not maintained in the low energy side because the pulse heights
of charge-sharing pixels could be effectively in the range of the low energy side.
This effect makes the line shape asymmetric, having non-Gaussian tails in the high and/or low energy
side as is seen in Fig.~\ref{fig:pat20_spec} left.

The values of $\sigma_\mathrm{gain}$ are obtained as
0.5\% and 0.7\% from the single- and all-pixel event spectra, respectively.
Both values are lower than the original value of 1.2\% in
Fig.~\ref{fig:gain_variation}, indicating that the pixel gain correction certainly
works, but not enough to eliminate all the variation. Especially, the larger
$\sigma_\mathrm{gain}$ of the all-pixel event spectrum suggests that degree of
inaccuracy in the pixel gain correction is greater in the low energy side.

\section{Summary}
We have presented a detailed study of the X-ray spectroscopic performance of
XRPIX8 with the optimized PDD structure. The study is performed on the
event-driven readout mode data taken with operation temperatures of $-60$~$^\circ$C.
The details of the post-DAQ processing are also presented. The energy resolutions in FWHM at 6.4~keV of
the single-pixel and all-pixel event spectra are $178~\pm1$~eV and $291~\pm1$~eV,
respectively. These are the best records in our XPRIX series.
We have discussed that the current spectroscopic performance is mainly limited by the gain non-linearity in the low
energy side due to excessive charge injection to the CSA circuit. Optimization of
the amount of the charge injection is expected to further improve the spectroscopic
performance of XRPIX, especially for the all-pixel event spectrum.

\section{Acknowledgments}
We acknowledge the valuable advice and the manufactures of XRPIXs by the personnel of LAPIS Semiconductor Co., Ltd.
This study was supported by JSPS KAKENHI Grant Numbers 21H01095, 23K20850 and by JST SPRING, Grant Number JPMJSP2105.
This study was also supported by the VLSI Design and Education Center (VDEC), the University of
Tokyo in collaboration with Cadence Design Systems, Inc., and Mentor Graphics, Inc.

\bibliography{mybibfile}

\begin{thebibliography}{10}
\expandafter\ifx\csname url\endcsname\relax
  \def\url#1{\texttt{#1}}\fi
\expandafter\ifx\csname urlprefix\endcsname\relax\def\urlprefix{URL }\fi
\expandafter\ifx\csname href\endcsname\relax
  \def\href#1#2{#2} \def\path#1{#1}\fi

\bibitem{ref:bigas}
M.~Bigas, et~al., Review of cmos image sensors, Microelectronics Journal 37~(5)
  (2006) 433--451.

\bibitem{ref:alarcon}
M.~R. Alarcon, et~al., Scientific cmos sensors in astronomy: Imx455 and imx411,
  Publications of the Astronomical Society of the Pacific 135~(1047) (2023)
  055001.

\bibitem{ref:Burke}
B.~E. Burke, et~al., Ccd soft x-ray imaging spectrometer for the asca
  satellite, IEEE Transactions on Nuclear Science 41~(1) (1994) 375--385.

\bibitem{ref:Garmire}
G.~P. Garmire, et~al., Advanced ccd imaging spectrometer (acis) instrument on
  the chandra x-ray observatory, in: Proceedings of SPIE 4851 (2003) 28--44.

\bibitem{ref:Turner}
M.~J.~L. Turner, et~al., The european photon imaging camera on xmm-newton: The
  mos cameras, Astronomy and Astrophysics 365~(1) (2001) L27--L35.

\bibitem{ref:Koyama}
K.~Koyama, et~al., X-ray imaging spectrometer (xis) on board suzaku,
  Publications of the Astronomical Society of Japan 59 (2007) S23--S33.

\bibitem{ref:Predehl}
P.~Predehl, et~al., The erosita x-ray telescope on srg, Astron. Astrophys. 647
  (2021) A1.

\bibitem{ref:Tanaka}
T.~Tanaka, et~al., Soft x-ray imager aboard {H}itomi ({ASTRO}-{H}), Journal of
  Astronomical Telescopes, Instruments, and Systems 4~(1) (2018).

\bibitem{ref:Nakajima}
H.~Nakajima, et~al., In-orbit performance of the soft x-ray imaging system
  aboard hitomi (astro-h), Publications of the Astronomical Society of Japan
  70~(2) (2018) 21.

\bibitem{ref:Mori}
K.~Mori, et~al., Status of xtend telescope onboard x-ray imaging and
  spectroscopy mission (xrism), in: Proceedings of SPIE 13093 (2024) 130931I.

\bibitem{ref:Bautz}
M.~Bautz, et~al., Fast, low-noise image sensor technology for strategic x-ray
  astrophysics missions, in: Proceedings of SPIE 13093 (2024) 130931Q.

\bibitem{ref:Meidinger}
N.~Meidinger, et~al., erosita camera array on the srg satellite, Journal of
  Astronomical Telescopes, Instruments, and Systems 7~(2) (2021) 025004.

\bibitem{ref:Eric}
E.~D. Miller, et~al., The high-speed x-ray camera on axis, in: UV, X-Ray, and
  Gamma-Ray Space Instrumentation for Astronomy XXIII, Vol. 12678, SPIE, 2023,
  p. 1267816.

\bibitem{ref:Qinyu}
W.~Q, et~al., Improving the x-ray energy resolution of a scientific cmos
  detector by pixel-level gain correction, Publications of the Astronomical
  Society of the Pacific 135~(1044) (2023) 025003.

\bibitem{ref:Tsuru}
T.~G. Tsuru, H.~Hayashi, K.~Tachibana, et~al., Kyoto's event-driven x-ray
  astronomy soi pixel sensor for the force mission, in: Proceedings of SPIE
  10709 (2018).

\bibitem{ref:Arai}
Y.~Arai, T.~Miyoshi, Y.~Unno, et~al., Development of {SOI} pixel process
  technology, Nuclear Instruments and Methods in Physics Research Section A:
  Accelerators, Spectrometers, Detectors and Associated Equipment 636 (2011)
  S31--S36.

\bibitem{ref:Takeda_1b}
A.~Takeda, Y.~Arai, S.~G. Ryu, et~al., Design and evaluation of an soi pixel
  sensor for trigger-driven x-ray readout, IEEE Transactions on Nuclear Science
  60~(2) (2013) 586--591.

\bibitem{ref:Kamehama}
H.~Kamehama, S.~Kawahito, S.~Shrestha, et~al., A low-noise x-ray astronomical
  silicon-on-insulator pixel detector using a pinned depleted diode structure,
  Sensors 18~(1) (2018) 27.

\bibitem{ref:Yukumoto}
M.~Yukumoto, K.~Mori, A.~Takeda, et~al., Design study and spectroscopic
  performance of soi pixel detector with a pinned depleted diode structure for
  x-ray astronomy, Nuclear Instruments and Methods in Physics Research Section
  A: Accelerators, Spectrometers, Detectors and Associated Equipment 1060
  (2024) 169033.

\bibitem{ref:Hagino_3b}
K.~Hagino, K.~Oono, K.~Negishi, et~al., Measurement of charge cloud size in
  x-ray soi pixel sensors, IEEE Transactions on Nuclear Science 66~(7) (2019)
  1897--1905.

\bibitem{ref:Ryu_1}
S.~G. Ryu, T.~G. Tsuru, S.~Nakashima, et~al., First performance evaluation of
  an x-ray soi pixel sensor for imaging spectroscopy and intra-pixel trigger,
  IEEE Transactions on Nuclear Science 58~(5) (2011) 2528--2536.

\bibitem{ref:janesick}
J.~R. Janesick, Scientific Charge-Coupled Devices, SPIE, 2001.

\bibitem{ref:Harada}
S.~Harada, T.~G. Tsuru, T.~Tanaka, et~al., Performance of the
  silicon-on-insulator pixel sensor for x-ray astronomy, xrpix6e, equipped with
  pinned depleted diode structure, Nuclear Instruments and Methods in Physics
  Research Section A: Accelerators, Spectrometers, Detectors and Associated
  Equipment 924 (2019) 468--472.

\bibitem{ref:Kodama}
R.~Kodama, T.~G. Tsuru, T.~Tanaka, et~al., Low-energy x-ray performance of
  {SOI} pixel sensors for astronomy, ”{XRPIX}”, Nuclear Instruments and
  Methods in Physics Research Section A: Accelerators, Spectrometers, Detectors
  and Associated Equipment 986 (2021) 164745.

\bibitem{ref:Mori2022}
K.~Mori, et~al., Xtend, the soft x-ray imaging telescope for the x-ray imaging
  and spectroscopy mission (xrism), in: Proceedings of SPIE 12181 (2022).

\bibitem{ref:Kotov}
I.~V. Kotov, et~al., Pair creation energy and fano factor of silicon measured
  at 185 k using x-rays, Nuclear Instruments and Methods in Physics Research
  Section A: Accelerators, Spectrometers, Detectors and Associated Equipment
  901 (2018) 126--132.

\end{thebibliography}

\end{document}